\documentstyle[12pt]{article}
\def\half{\frac{1}{2}}

\def\dX2{d\vec{x}\cdot d\vec{x}}
\newcommand{\be}{\begin{equation}}
\newcommand{\ee}{\end{equation}}
\newcount\sectionnumber
\def\sect
{\global\equationnumber=0
\global\advance\sectionnumber by 1
\the\sectionnumber . }
\newcount\equationnumber
\def   \num
{\eqno{\global\advance\equationnumber by 1
\left(\the\sectionnumber .\the\equationnumber \right)}}

\begin{document}
\pagenumbering{roman}

\title{
TRAVERSABLE WORMHOLES IN (2+1) AND (3+1) DIMENSIONS WITH
A COSMOLOGICAL CONSTANT}
\author{M. S. R. Delgaty \\
Physics Department,\\
Queen's University,\\
Kingston, Ontario, Canada\\
\vspace{10pt}
K7L 3N6\\
R. B. Mann\\
Physics Department,\\
University of Waterloo,\\
Waterloo, Ontario, Canada\\
N2L 3G1}
\date{}
\maketitle

\begin{abstract}

Macroscopic traversable wormhole solutions to Einstein's field equations in
$(2+1)$ and $(3+1)$ dimensions with a cosmological constant are
investigated.   Ensuring traversability severely constrains the material
used to generate the wormhole's spacetime curvature.  Although the presence
of a cosmological constant modifies to some extent the type of matter
permitted (for example it is possible to have a positive energy density for
the material threading the throat of the wormhole in $(2+1)$ dimensions),
the material must still be ``exotic'', that is matter with a larger radial
tension than total mass-energy density multiplied by $c^2$.  Two specific
solutions are applied to the general cases and a partial stability analysis
of a $(2+1)$ dimensional solution is explored.

\vspace{10pt}
WATPHYS TH-93/06

\end{abstract}

\pagenumbering{arabic}

%
%%%%%%%%%%%%%%%%%%%%%%%%%%%%%%%%%%%%%%%%%%%%%%%%%%%%%%%%%%%%%%%%%%%%%%%%%%%%
%
\section{Introduction}
\bigskip

Wormholes are tunnels in the geometry of space and time that connect two
separate and distinct regions of spacetime.  These regions may either be
part of the same universe or be regions of two different universes.
Although such objects were long known to be solutions to Einstein's
equations, early work on macroscopic wormholes  (those considered large
enough for interstellar travel) led to the conclusion that at best they
were either unstable or hidden behind event horizons, so as not to be
traversable by living entities.  For this reason wormholes have only
intermittently been studied, even though they predate black holes as
objects of interest in relativity.

Recently, a renaissance in the study of wormholes has taken place,
instigated by a closer investigation of conditions necessary to ensure
their traversability \cite{kn:MorThorne}.  This interest is motivated in
part by the possibility that quantum
gravity might permit the formation of ``exotic'' material  ({\it ie}.
matter which violates the positive energy condition) necessary to construct
traversable wormholes. Furthermore, processes involving quantum gravity are
important in both early universe cosmology \cite{kn:Coleman} and the final
stages of black hole evaporation \cite{kn:Page93}.

The main focus of this paper is traversable macroscopic wormholes in
$(2+1)$ and $(3+1)$ dimensional spacetimes with a cosmological constant.
Previous investigations into wormholes have not considered this case
\cite{kn:MorThorne,kn:PerryMann}, although there has been some
investigation of the Wheeler-DeWitt equation for wormholes in spacetimes
with a cosmological constant \cite{kn:Kim}. As the cosmological constant
can be interpreted as the energy density of the vacuum one might expect it
to modify the form of the exotic matter required. As we shall see, this
does in fact occur, although not in such a manner as to avoid violation of
the weak-energy condition. A positive constant will produce an expansion
term that counteracts gravity, while a negative constant will aid
gravitational collapse of the wormhole.  We will consider the case of a
positive constant ({\it ie}. anti-de Sitter spacetime)
in this paper unless otherwise stated.

Proceeding as in ref. \cite{kn:PerryMann}, we consider what constraints are
required for the wormhole to be traversable in $(2+1)$ and $(3+1)$
dimensions.   Section III presents the needed tensors and solves Einstein's
field equations.  Through the use of embedding, inequality restrictions on
the type of matter that is needed to generate the wormhole's spacetime
curvature are presented, modifying the results of ref. \cite{kn:PerryMann}.
Section 4 discusses the mathematical equivalents of the traversability
criteria.  These criteria are utilized in Section 5 where the zero-tidal
force solutions for both $(2+1)$ and $(3+1)$ dimensions.  Additionally the
``Junction Condition'' Formalism solution with stability analysis for the
simplest case is investigated in $(2+1)$ dimensions. %

%%%%%%%%%%%%%%%%%%%%%%%%%%%%%%%%%%%%%%%%%%%%%%%%%%%%%%%%%%%%%%%%%%%%%%%%%%%%
% \section{Wormhole Constraints}\label{sec2} \bigskip

Wormhole solutions that connect regions in the same universe or regions
in two different universes are the same, but the topology for each case,
which is not constrained by the field equations, is different.  The
important facet of the wormhole is the throat which is a finite spatial
region between the wormhole `mouths', at which the embedding surface
(discussed below) will be vertical.  This exists at a (possibly large)
set of circles (spheres) of minimum radius $r$, where $r$ is
the radial coordinate, producing a throat that travellers could use
as their bridge.

We employ (in the context of the Einstein field equations)
the traversability properties of ref. \cite{kn:MorThorne} in
$(3+1)$ dimensions and of \cite{kn:PerryMann} in $(2+1)$ dimensions.
Briefly, these are
\medskip
\begin{itemize}
\item[(1)] The metric should be spherically (radially)
symmetric and static (i.e. no time dependence).
\item[(2)] By definition, the solution must have a throat which connects
two regions of spacetime.
\item[(3)] There should be no horizon so as to permit two way travel.
\item[(4)] The tidal gravitational forces experienced by a
traveller must be reasonably small (eg. approximately one Earth gravity).
\item[(5)] The time to traverse the wormhole must be reasonably short (eg.
one year) as measured by both the traveller and any observers who wait on
the outside of the wormhole.
\end{itemize}

Criteria (1) through (3) can be thought of as basic to constructing
wormholes, while properties (4) and (5) are usability criteria.  Property
(1) is not a requirement for the wormhole solution, but it greatly
simplifies the calculations.  It is still possible that the wormhole might
be unstable to radial or non-radial perturbations, a subject which we will
investigate to a certain extent in section 5.

We choose the following metrics for the $(2+1)$ and $(3+1)$ cases
respectively:
\be
ds^2 = -e^{2{\Phi}(r)}c^{2} dt^2 + \frac{dr^2}{{\Lambda}{r^2}-{M(r)}} +
r^2 d{\phi}^2, \label{eq:metric1}
\ee
\be
ds^2 = -e^{2{\Phi}(r)}c^{2} dt^2 + \frac{dr^2}{\frac{\Lambda r^2}{3}-
\frac{M(r)}{r}} + r^2(d{\theta}^2 + sin^2{\theta} d{\phi}^2),
\label{eq:metric2}
\ee
where $\Lambda $ is the cosmological constant with units of
$\mbox{cm}^{-2}$ and $M$ and $\Phi$ are arbitrary functions of the radial
coordinate $r$ (unless otherwise specified $M \equiv M(r)$ and $\Phi \equiv
\Phi (r)$).  The $(3+1)$ dimensional metric is similar to the DeSitter
metric with the constant incorporated into the function $M$ to simplify
calculations and make it easier to identify the physics involved.  The
shape of the wormhole is specifically chosen by the designer and is
modified by the ``shape function'' $M$ \cite{kn:MorThorne}.  $\Phi $
determines the gravitational redshift and will be referred to as the
``redshift'' function.  We see that the metrics are respectively cirularly
and spherically symmetric, and static.  Choosing $\Phi (r)$ finite ensures
that there is no horizon (the time component does not vanish), and note
that there is in fact a throat when $\Lambda r^2$ equals $M$, and when
$\Lambda r^2/3$ equals $M/r$ respectively.
%
%%%%%%%%%%%%%%%%%%%%%%%%%%%%%%%%%%%%%%%%%%%%%%%%%%%%%%%%
%
\section{Stress-Energy Constraints}
Consider metric (\ref{eq:metric1}) above.  By introducing a set of
orthonormal (hatted) basis vectors (that is the reference frame
of a set of observers who remain always at rest in the coordinate system), the
mathematics and physical interpretations become greatly simplified.  The new
basis vectors are,
\be
{\bf e}_{\hat{t}}=e^{-\Phi}{\bf e}_{t} \mbox{,\ \ \ } {\bf e}_{\hat{r}}=
(\Lambda {r^2}-M)^{1/2}{\bf e}_{r}
\mbox{,\ \ \ } {\bf e}_{\hat{\phi}}=r^{-1}{\bf e}_{\phi},
\label{eq:wormframe1}
\ee
where ${\bf e}_{t} = c^{-1}\partial / \partial t, {\bf e}_{r} = \partial /
\partial r, $and ${\bf e}_{\phi}=\partial / \partial\phi$.  This gives
$g_{\hat{\alpha}\hat{\beta}}=\eta_{\hat{\alpha}\hat{\beta}}$, and we find
the following Riemann tensor components:
\begin{eqnarray}
R_{\hat{t}\hat{r}\hat{t}\hat{r}} & = & e^{-2\Phi}(\Lambda r^2 -M)R_{trtr}
\nonumber \\ & = & (\Lambda r^2 -M)\left[\Phi ''+\frac{2\Lambda r-M'}
{2(\Lambda {r^2}-M)}\Phi' +(\Phi ')^2\right], \label{eq:rtrt}
\end{eqnarray}
\be
R_{\hat{t}\hat{\phi}\hat{t}\hat{\phi}}=e^{-2\Phi}r^{-2}R_{t\phi t\phi}=
(\Lambda {r^2}-M)\Phi '/r, \label{eq:rptp}
\ee
\be
R_{\hat{r}\hat{\phi}\hat{r}\hat{\phi}}=r^{-2}(\Lambda r^2 -M)R_{r\phi r\phi}=
\frac{M'-2\Lambda r}{2r}, \label{eq:rprp}
\ee
where the prime denotes a derivative with respect to $r$.
The Einstein tensor $G_{\hat{\alpha}\hat{\beta}}\equiv R_{\hat{\alpha}
\hat{\beta}}-{\textstyle\frac{1}{2}} g_{\hat{\alpha}\hat{\beta}}R$
has the following non-zero components:
\be
G_{\hat{t}\hat{t}} = \frac{M'}{2r}-\Lambda,  \label{eq:gtt1}
\ee
\be
G_{\hat{r}\hat{r}}=\frac{(\Lambda {r^2}-M)\Phi'}{r},  \label{eq:grr1}
\ee
\be
G_{\hat{\phi}\hat{\phi}}={\left(\Lambda {r^2}-M\right)}{\left[\Phi ''+
\frac{2\Lambda r-M'}{2(\Lambda r^2-M)}\Phi '+(\Phi ')^2\right]}.
\label{eq:gpp1}
\ee
Using the same method on (\ref{eq:metric2}), we get the following for the
$(3+1)$
dimensional case:
\be
G_{\hat{t}\hat{t}} = \frac{1+M' - \Lambda r^2}{r^2},  \label{eq:gtt2}
\ee
\be
G_{\hat{r}\hat{r}}=\frac{2\Phi '}{r^2}\left( \frac{\Lambda r^3}{3}-M \right) +
\frac{\Lambda r^3 -3M-3r}{3r^3},\label{eq:grr2}
\ee
\begin{eqnarray}
G_{\hat{\theta}\hat{\theta}}=G_{\hat{\phi}\hat{\phi}}&=& \left(\frac{\Lambda
r^3}{3}-\frac{M}{r}\right)\left[\Phi '' + \Phi' \left(\frac{2\Lambda r^3 -3rM'
+3M}{2r(\Lambda r^3 -3M)}+\frac{1}{r} \right) \right. \nonumber \\
                                                     & & \left. + (\Phi ')^2
+\frac{2\Lambda r^3-3rM'+3M}{2r^2 (\Lambda r^3-3M)}\right].
\label{eq:gpp2}
\end{eqnarray}

The field equations
\be
G_{\hat{\alpha}\hat{\beta}}-\Lambda g_{\hat{\alpha}\hat{\beta}}
=\kappa T_{\hat{\alpha}\hat{\beta}},
\ee
where $\kappa\equiv 8\pi Gc^{-4}$ imply that
the stress-energy tensor $T_{\hat{\alpha}\hat{\beta}}$ must be of the same
form as the Einstein tensor.  Hence the only non-zero components are
\be
T_{\hat{t}\hat{t}}=\rho (r)c^2,\ T_{\hat{r}\hat{r}}=-\tau (r),\
T_{\hat{\theta}\hat{\theta}}=p(r)=T_{\hat{\phi}\hat{\phi}}, \label{eq:set}
\ee
where $\rho (r)$ is the total density of mass-energy that the static observers
measure in units of $\mbox{g/cm}^2$ ($\mbox{g/cm}^3$ for the $(3+1)$ case);
$\tau (r)$ is the radial tension per unit area that they measure [it is
the negative of the radial pressure and has units $\mbox{dyn/cm}$]; and $p(r)$
is the lateral pressure that they measure
in directions orthogonal to the radial direction.

Solving these equations for $\rho (r)$, $\tau (r)$, and $p(r)$ we get:
\be
\rho (r)=\frac{c^2}{8\pi G}\left(\frac{M'}{2r}\right), \label{eq:efer}
\ee
\be
\tau (r)=\frac{c^4}{8\pi G}\left(\Lambda -\frac{\Phi'}{r}(\Lambda
{r^2}-M)\right), \label{eq:efet}
\ee
and
\be
p(r) = {\frac{c^4}{8\pi G}}\left[ (\Lambda {r^2}-M)\left({\Phi}''
+\frac{2\Lambda r-M'}{2(\Lambda {r^2}-M)}\Phi '+({\Phi}')^2\right)-\Lambda
\right], \label{eq:efep}
\ee
for the $(2+1)$ case and
\be
\rho (r)=\frac{c^2}{8\pi Gr^2}(1+M'), \label{eq:efer2}
\ee
\be
\tau (r)=\frac{c^4}{8\pi Gr^2}\left[\left(\frac{2\Lambda r^2}{3}+\frac{M}{r} +1
\right) - 2\Phi 'r \left(\frac{\Lambda r^2}{3} - \frac{M}{r}\right)
\right], \label{eq:efet2}
\ee
and
\begin{eqnarray}
p(r)&=& \frac{c^4}{8\pi G}\left[\left(\frac{\Lambda
r^2}{3}-\frac{M}{r}\right)\left({\Phi}'' + \left(\frac{2\Lambda r^3-3rM'
+3M}{2r(\Lambda r^3-3M)} +\frac{1}{r}\right)\Phi ' \right. \right. \nonumber \\
    & & \left. \left. + (\Phi ')^2 + \frac{2\Lambda r^3-3rM' +3M}{2r^2(\Lambda
r^3-3M)}\right)-\Lambda\right], \label{eq:efep2}
\end{eqnarray}
for the $(3+1)$ case.
The typical way to solve these equations would be to assume a particular type
of material, and then derive equations of state based on that material.  This
would then give the three equations above with two equations of state (one for
$\tau (\rho)$, and one for $p(\rho)$) for a total of five equations with five
unknown functions of $r$.  Since we are concerned with traversable wormholes
with certain properties, the solutions come by tailoring the choices of
$\Phi (r)$ and $M(r)$ for a ``nice'' wormhole.  In this fashion we have the
three equations above as functions of $r$.  The requirement of this method is
that the wormhole builders must in some way obtain the material with the
appropriate stress-energy tensor so calculated.  The above process is applied
in Section 4 to the zero tidal-force solution.

Consider next the embedding of the spatial geometry of the wormhole into a
flat space of one higher dimension. First, take a ``slice'' of the metric
at constant time (at $t=t_0=$constant, $dt=0$), and embed this slice into
the usual flat Lorentzian space:
\be
ds^2 = dz^2  + dr^2 + r^2 d{\phi}^2, \label{eq:zmetric}
\ee
and for the $(3+1)$ case
\be
ds^2 = dz^2 + dr^2 + r^2 (d{\theta}^2 +
sin^2{\theta} d{\phi}^2). \label{eq:zmetric2}
\ee
The geometry of (\ref{eq:zmetric}), and (\ref{eq:zmetric2}) implies $z=z(r)$
due to the circular and spherical symmetry.  Hence we can solve for $z(r)$ by
setting the metrics for each case equal to get,
\be
\frac{dz}{dr} = \pm \left(\frac{1- \Lambda r^2 + M}{\Lambda r^2 -M}
\right)^{1/2}, \label{eq:dzdr}
\ee
for the $(2+1)$ case and for the $(3+1)$ case:
\be
\frac{dz}{dr} = \pm \left(\frac{1- \frac{\Lambda r^2}{3}+\frac{M}{r}}
{\frac{\Lambda r^2}{3}-\frac{M}{r}}\right)^{1/2}. \label{eq:dzdr2}
\ee
Solving these equations will yield the functional form
$z(r)$ of the two dimensional embedding surface.

Note that embedding in flat space is possible only if $\frac{dz}{dr}$
is real, implying
\be
M_{(2+1)} \leq \Lambda r^2 \leq M_{(2+1)} +1~~~ \mbox{and} ~~~ M_{(3+1)}
\leq \Lambda r^3 /3 \leq M_{(3+1)} + 1
\label{eq:throat}
\ee
for the respective cases. This is equivalent to requiring that $|dr/dl|
\leq 1$,  where
\be
dl_{2+1} = \frac{\pm dr}{\sqrt{\Lambda r^2 -M}},
\qquad \qquad
dl_{3+1} =\frac{\pm dr}{\sqrt{\frac{\Lambda r^2}{3}-\frac{M}{r}}}.
\label{eq:radial4}
\ee
The coordinate $l$ represents the proper radial distance from the
wormhole and ranges from zero to infinity.

Geometrical visualization is not the only use of embedding.  Since the
throat is a minimum radius from the z-axis and we know that at some point
away from the wormhole the space is essentially flat, (this is known since
a positive $\Lambda$ acts as a repulsive force at large distances, and must
at some radius balance gravity) we know that the embedding surface flares
outward.  For this to be true the matter distribution must be small
relative to $\Lambda ^{-1/2}$ so that the spacetime far from the wormhole
is not greatly affected by the matter.  Additionally, the wormhole itself
must also be small compared to the characteristic length so that it is not
locally deSitter and hence indistinguishable from the surrounding
Lorentzian space.  Assuming these properties we take ${d^2}r/d{z^2} > 0$
and $r(z)$ to be a minimum at the throat.  Inverting (\ref{eq:dzdr}) and
(\ref{eq:dzdr2}) and taking the derivative respectively gives the following
useful conditions:
\be
M' < 2 \Lambda r,~\mbox{and}~~\frac{M'}{r} <
\frac{2\Lambda r}{3} +\frac{M}{r^2}.  \label{eq:flairy}
\ee
By applying the
first inequality of (\ref{eq:throat}) and (\ref{eq:flairy}) we get
\be \rho(r) < \frac{c^2\Lambda}{8\pi G}, \label{eq:rhoequ}
\ee
in $(2+1)$ dimensions. Note that when $\Lambda=0$ \cite{kn:PerryMann}
$\rho < 0$; we see that the addition of a positive $\Lambda$ increases the
energy density for that case.  In $(3+1)$ dimensions we get
\be
\rho (r) < \frac{c^2}{8\pi Gr^2}\left(1+\frac{4\Lambda r^2}{3}\right).
\ee
in agreement the $\Lambda=0$ case in that both allow for the possibility of
a positive mass-energy for a positive $\Lambda$ \cite{kn:MorThorne}.

Another physical aspect of the wormhole can be investigated by forming
the dimensionless ratio \cite{kn:MorThorne}
\be
\zeta\equiv \frac{\tau - \rho c^2}{|\rho c^2|}.
\ee
For regular matter (that respects the weak energy condition)
this ratio is negative; using conditions (\ref{eq:flairy})
and (\ref{eq:throat}) at the
throat (where the restrictions are the most severe)
along with the fact that $\Phi'$ is finite, we find for both
cases that
\be
\zeta_0\equiv \frac{\tau_0 - \rho_0 c^2}{|\rho_0 c^2|} >0.
\ee
Here the subscript $0$ signifies that the quantity is evaluated at the
minimum radius.  That is $r=r_0 \neq 0 \Rightarrow \Lambda r_{0}^{2} -M_0 =0$
for (\ref{eq:dzdr}), and $\Lambda r_{0}^{3} = 3M_0$ for equation
(\ref{eq:dzdr2}), where $M_0 \equiv M(r_0)$.  As in ref. \cite{kn:MorThorne}
we shall call material with the property $\tau > \rho c^2>0$,
``exotic''.  This terminology arises because
an observer moving through the throat with sufficiently large velocity
will necessarily see a negative mass-energy density.

A negative density of mass-energy implies a violation of the ``weak energy
condition'' \cite{kn:MorThorne}.  This in itself gives no basis for
immediately discarding solutions as the weak energy condition has been
experimentally shown to be false in $(3+1)$ dimensions, the Casimir effect
being the perhaps the best known. Whether or not exotic matter can be
formed in macroscopic quantities is still an open question.

%
%%%%%%%%%%%%%%%%%%%%%%%%%%%%%%%%%%%%%%%%%%%%%%%%%%%%%%%%%%%%%%%%%%%%%%%%%%%%
%
\section{Traversability Criteria}

As discussed in Section 1, in order for the wormhole to be useful for
theoretical $(2+1)$ dimensional beings it must take a reasonable time
to traverse the wormwhole as seen by both the traveller and those waiting
outside the wormhole.  Additionally, the tidal forces experienced by the
traveller must not be too great (say the equivalent to one Earth gravity).

These criteria are analogous to those in references \cite{kn:MorThorne} and
\cite{kn:PerryMann}.  Consider a wormhole that joins two distinct universes.
Assume that the traveller starts at location $l=-l_1$ in a lower universe
(pictorially speaking) at rest and ends at $l=l_2$ in an upper universe at
rest.  As in special relativity, denote $\gamma\equiv(1 - (v/c)^2)^{-1/2}$.
Noting that $dl$ is the distance travelled, $dr$ is the radius travelled,
$dt$ is the coordinate time lapse, and $d\tau_{\mbox{\tiny T}}$ is the proper
time lapse as seen by the observer, we have for the $(2+1)$ case
\be
v=\frac{dl}{e^\Phi dt}= \mp \frac{dr}{\sqrt{\Lambda r^2 - M}e^\Phi dt},~\mbox
{and}~~v\gamma =\frac{dl}{d\tau_{\mbox{\tiny T}}}=\mp \frac{dr}
{\sqrt{\Lambda r^2-M}d\tau_{\mbox{\tiny T}}}, \label{eq:radial2}
\ee
and for the $(3+1)$ case
\be
v=\mp \frac{dr}{\sqrt{\frac{\Lambda r^2}{3} -\frac{M}{r}}e^\Phi dt},~\mbox
{and}~~v\gamma =\mp \frac{dr}
{\sqrt{\frac{\Lambda r^2}{3}-\frac{M}{r}}d\tau_{\mbox{\tiny T}}},
\label{eq:radial3}
\ee
where the ($-$) sign refers to the first half of the trip from station
$l_1$ to the wormhole and the ($+$) sign refers to the second half trip from
the
wormhole to station $l_2$.

It is desirable that the effects of the wormhole are small at the
stations.  This can be effectively obtained by specifying: i)
the positions of the stations such that the geometry is essentially (to
less than one percent) the same as space which is not affected by the wormhole
($M_{(2+1)} \ll \Lambda r^2$ or $M_{(3+1)}/r \ll \Lambda r^2/3$); ii) the
gravitational
red-shift of signals sent off to
infinity from the stations is small ($|\Phi | \ll 1$); iii) the ``acceleration
of gravity'' as measured at the stations is no more than 1 Earth gravity
$\equiv g_{\oplus}(=980 \mbox{cm/s}^2)$ \cite{kn:MorThorne};
\be
|g| \approx |-\Phi 'c^2| \leq g_{\oplus}
\ee

Now, the time criteria can be satisfied by the equations
\be
\Delta\tau_{\mbox{\tiny T}}= \int^{l_2}_{-l_1} \frac{dl}{v\gamma} \leq 1
\mbox{ yr., for the traveler and,} \label{eq:ptime}
\ee
\be
\Delta t = \int^{l_2}_{-l_1} \frac{dl}{ve^\Phi}\leq 1 \mbox{ yr., for outside
observers.} \label{eq:ctime}
\ee
It is also reasonable to assume that the maximum acceleration the traveller
can experience is about 1 Earth gravity.  This gives the restriction
\be
\left| e^{-\Phi}\frac{d(\gamma e^\Phi)}{dl}\right| \leq \frac{g_\oplus}{c^2}
\simeq \frac{1}{0.97}~\frac{1}{\mbox{l.~yr.}} \label{eq:const3}
\ee
In general, the tidal acceleration is given by
\be
\Delta a^{{\hat \alpha}'} = - c^2 R^{{\hat \alpha}'}{}_{{{\hat \beta}'}
{{\hat \gamma}'}{{\hat \sigma}'}} u^{{\hat \beta}'} \xi^{{\hat \gamma}'}
u^{{\hat \sigma}'}. \label{eq:tide}
\ee
where $\xi^\mu$ is a spacelike vector which
denotes the  separation of two parts of the traveller's body and
$u^\mu$ is the four-velocity.  Note that
$u^{{\hat \alpha}'}=\delta^{{\hat \alpha}'}_{{\hat 0}'}$ and
$\xi^{{\hat 0}'}=0$ in the travellers frame.  Additionally since the Riemann
tensor is anti-symmetric in its first two indices, the tidal accelerations
are purely spatial with components,
\be
\Delta a^{{\hat j}'} = - c^2 R^{{\hat j}'}{}_{{{\hat 0}'}
{{\hat k}'}{{\hat 0}'}} \xi^{{\hat k}'}. \label{eq:tidey}
\ee

Taking the size of a human to be $|{\bf \xi }|\sim 2$m and
$|\Delta {\bf a}| \leq g_{\oplus}$ for $\xi$
oriented along any spatial direction in the travelers frame, then the Riemann
tensor components are constrained to obey,
\begin{eqnarray}
\left|R_{{{\hat 1}'}{{\hat 0}'}{{\hat 1}'}{{\hat 0}'}}\right| & = &
\left| \left(\Lambda r^2-M\right)\left(\Phi '' + \frac{2\Lambda r-M'}
{2(\Lambda r^2-M)}\Phi' + (\Phi ')^2 \right) \right| \nonumber \\
& \leq & \frac{g_\oplus}{c^2\cdot 2\,\mbox{m}}\simeq\frac{1}{(10^{10}
\mbox{cm})^2}, \label{eq:const1}
\end{eqnarray}
and
\begin{eqnarray}
\left|R_{{{\hat 2}'}{{\hat 0}'}{{\hat 2}'}{{\hat 0}'}}\right| & = &
\left| \frac{\gamma ^2}{2r}\left[\left(\frac{v}{c}\right)^2
\left(M'-2\Lambda r\right) - 2\Phi '(\Lambda r^2-M)\right] \right| \nonumber \\
& \leq & \frac{g_\oplus}{c^2\cdot 2\,\mbox{m}}\simeq\frac{1}{(10^{10}
\mbox{cm})^2}, \label{eq:const2}
\end{eqnarray}
for the $(2+1)$ case and for the $(3+1)$ case
\begin{eqnarray}
\left|R_{{{\hat 1}'}{{\hat 0}'}{{\hat 1}'}{{\hat 0}'}}\right| & = &
\left| \left(\frac{\Lambda r^2}{3}-\frac{M}{r}\right)\left(\Phi '' +
\frac{2\Lambda r^3-3rM' +3M}{2r(\Lambda r^3 -3M)}\Phi' +(\Phi
')^2\right)\right|
\nonumber \\
& \leq & \frac{g_\oplus}{c^2\cdot 2\,\mbox{m}}\simeq\frac{1}{(10^{10}
\mbox{cm})^2}, \label{eq:const12}
\end{eqnarray}
and
\begin{eqnarray}
\left|R_{{{\hat 2}'}{{\hat 0}'}{{\hat 2}'}{{\hat 0}'}}\right| & = &
\left| \frac{\gamma ^2}{2r}\left[\left(\frac{v}{c}\right)^2
\left(\frac{2 \Lambda r}{3}-\frac{M'}{r}+\frac{M}{r^2}\right) +
2\Phi '\left(\frac{\Lambda r^2}{3}-\frac{M}{r}\right)\right] \right| \nonumber
\\
& \leq & \frac{g_\oplus}{c^2\cdot 2\,\mbox{m}}\simeq\frac{1}{(10^{10}
\mbox{cm})^2}. \label{eq:const22}
\end{eqnarray}

Equations (\ref{eq:const1}) and (\ref{eq:const12}) represent the radial tidal
force constraint and can be regarded as constraining the function $\Phi (r)$
while equations (\ref{eq:const2}) and (\ref{eq:const22}) represent the lateral
tidal force and restrict the speed $v$ of the traveller while crossing
the wormhole.

%
%%%%%%%%%%%%%%%%%%%%%%%%%%%%%%%%%%%%%%%%%%%%%%%%%%%%%%%%%%%%%%%%%%%%%%%%%%%%
%
\section{Wormhole Solutions}

The first solution presented results when $\Phi =0$ everywhere.  This solution
is called the zero tidal force because a stationary observer $(v=0)$ will not
experience any tidal forces [c.f. Eqs. (\ref{eq:const1}), (\ref{eq:const2}) and
(\ref{eq:const12}), (\ref{eq:const22})].  Consider the following choices which
satisfy the conditions of Section 1, equations (\ref{eq:throat}),
(\ref{eq:flairy}), and (\ref{eq:rhoequ}), and allow for fairly simple
integral equations;
\be
\Phi = 0,~~~M_{(2+1)} = -k\Lambda r^2+k/100,~~~M_{(3+1)}=-k \Lambda r^3/3 +
kr/100 \label{eq:slb}
\ee
where $k$ is some constant greater than zero so that
the signature of $dr$ is maintained and we have a throat.  Substitution
of (\ref{eq:slb}a,b) into
(\ref{eq:efer}), (\ref{eq:efet}), and (\ref{eq:efep}), and (\ref{eq:slb}a,c)
into (\ref{eq:efer2}), (\ref{eq:efet2}), and (\ref{eq:efep2}) gives:
\be
\frac{c^2 \rho (r)}{k}= -\tau (r)=p(r)=\frac{-c^4\Lambda}{8\pi G},
\label{eq:sefer}
\ee
for the $(2+1)$ case and for the $(3+1)$ case:
\be
\rho (r)= \frac{c^2}{8\pi Gr^2}(1+k/100 - k \Lambda r^2), \label{eq:sefer2}
\ee
\be
\tau (r)=\frac{c^4}{8\pi Gr^2}\left(\frac{ \Lambda r^2}{3}(2-k)
+\frac{k}{100} +1 \right),  \label{eq:sefet2}
\ee
\be
p(r)=\frac{\Lambda c^4}{24\pi G}(k-2). \label{eq:sefep}
\ee

Equation (\ref{eq:sefer}) shows that the energy-density of exotic matter
depends on the value
of $\Lambda$ and is constant throughout the universe.  This makes the solution
unrealistic.  A possible way around this problem would be to create the
wormhole within a relativistic vacuum bubble as described in \cite{kn:AKMS};
the interior of the bubble would have a non-zero $\Lambda$, but the exterior
could have $\Lambda = 0$. Such a bubble would contain the exotic matter in
some finite volume of spacetime.  Travellers wishing to go through the
wormhole would have to pass through the discontinuity at the surface of the
bubble.  Equation (\ref{eq:sefer2}) shows that the exotic matter extends out
to infinity and $\rho(r)$ approaches a constant as $r \rightarrow
\infty$.  A vacuum bubble can only partially help in this case.  We would
still need a very large amount of exotic material, extending out to infinity,
to create such a wormhole.  The only other way around this is too have a
radial cutoff of the stress-energy (see \cite{kn:MorThorne}).

Integrating (\ref{eq:radial4}) for these solutions
yields the proper radial distaces as functions of $r$:
\be
l(r)= \mp \frac{1}{\sqrt{b}}ln\left[\frac{r + \sqrt{r^2-\frac{k}{100b}}}
{\sqrt{\frac{k}{100b}}}\right], \label{eq:leqn}
\ee
where $b_{(2+1)}=\Lambda (1+k)$, and $b_{(3+1)}=\Lambda (1+k)/3$.  The
two stations are located at a distance
$\Lambda r^2 - M \approx 1$ for the $(2+1)$ case and $\Lambda r^2/3-M/r
\approx 1$ for the $(3+1)$ case.  Taking this to be true to $1\%$
gives for both cases $r \approx \sqrt{(100+k)/(100b)}$.  Note that the
stations are located at the maximum value of $r$ that still allows embedding.
In order to use these solutions we mest therefore have a method of cutting
off larger $r$ values (see above paragraph).  These values of $r_0$ and $r$
increasingly conform to the requirement that
they be smaller than the characteristic length as $k$ increases.

Assuming that $(v/c) \ll 1$ gives $\gamma \approx 1$, with (\ref{eq:const2})
and (\ref{eq:const22}) we get:
\be
v \leq \frac{1}{\sqrt{b}}~~\mbox{ms}^{-1}.
\ee
where $b$ is defined for each case as before.
Hence (\ref{eq:ptime}) and (\ref{eq:ctime}) show us that for $\gamma \approx
1$;
\be
\Delta \tau_{\mbox{\tiny T}}\approx \Delta t \approx \int^{l_2}_{-l_1}
\frac{dl}{v} \simeq 2ln\left[\frac{10+\sqrt{100+k}}{\sqrt{k}}\right]~~\mbox{s},
\ee
These times depend only on the value of the constant $k$.  Large $k$ values
are better as they keep the wormhole smaller than the characteristic length,
but also reduce the proper time ($t \rightarrow 0$ as $k \rightarrow
\infty$).  The advantages of such a wormhole for human use is evident.

The next solution limits the exotic material to a circle of radius
$a$ around the wormhole.  Here we use the ``Junction Condition'' or
``Boundary Layer'' formalism \cite{kn:Visser,kn:Blau,kn:Visser2,kn:MTW}.
This solution will only be attempted for the $(2+1)$ dimensional case.  The
$(3+1)$ dimensional case can be done in an identical manner if needed.  The
model is constructed by surgically grafting the usual spacetime of
(\ref{eq:metric1}) between two identical (by oppositely directed)
spacetimes of the form
\be
ds^2 = -(\Lambda r^2-M_{+})c^2 dt^2 + \frac{dr^2}{\Lambda r^2-M_{+}} +
r^2 d{\phi}^2, \label{eq:jfup}
\ee
and,
\be
ds^2 = -(\Lambda r^2-M_{-})c^2 dt^2 + \frac{dr^2}{\Lambda r^2-M_{-}} +
r^2 d{\phi}^2. \label{eq:jfdown}
\ee
Note that $M_{+}$, and $M_{-}$ are constants, not functions.  By symmetry,
and the matching of the metrics for a continuous solution, it should be evident
that $M_{+} = M_{-} = M$.  Although the value of $M$ will be equivalent to
that of $M(r=a)$ in equation (\ref{eq:metric1}), we will not let $M(a)=M$ as
the derivatives will most likely not be equal.  This solution is a little
different than previous solutions as we now have a three part sandwich instead
of typical solutions which have simply two oppositely directed identical
spacetimes.

The two outer solutions have a zero stress-energy,
while the two boundary layers will both have a non-zero stress-energy.  The
magnitude of this stress-energy can be calculated in terms of the second
fundamental forms (equivalent to the extrinsic curvature tensor components)
at the boundaries.  For the following general formulas see ref.
\cite{kn:Visser}.  Adopting Riemann normal coordinates
at the junctions: $\eta$ denotes the coordinate normal to the junction with
$\eta$ positive in the manifold described by (\ref{eq:jfup}) and negative in
the manifold described by (\ref{eq:metric1}); and $x^{\mu}=(\tau,\phi,
\eta)$.  The second fundamental forms are given by:
\be
K^{i}{}_{j}^{\pm}=\half \left.g^{ik}\frac{\partial g_{kj}}{\partial \eta}
\right|_{\eta=\pm 0} =\half \left.\frac{\partial r}{\partial \eta}\right|_{r=a}
\left. g^{ik}\frac{\partial g_{kj}}{\partial r} \right|_{r=a} .
\label{eq:twoform}
\ee

The discontinuity in the second forms is then given by,
\be
{\cal K}_{ij}\equiv K^{+}_{ij}-K^{-}_{ij}. \label{eq:disc}
\ee
Conservation of stress-energy constrains the line stress-energy so
that the only non-zero components are $S^i_j$ with $S^{\eta\eta}=
0=S^{\eta i}$.  Additionally, the Einstein field equations lead to
\be
S^i{}_j=-\frac{c^4}{8\pi G}({\cal K}^{i}{}_{j}-
\delta^{i}_{j}{\cal K}^{k}{}_{k}).     \label{eq:linese}
\ee

Circular symmetry allows us to write
\be
{\cal K}^{i}{}_{j} = \left(
\begin{array}{rr}

        {\cal K}^{\tau}{}_{\tau} &            0  \\
                    0            & {\cal K}^{\theta}{}_{\theta}

\end{array} \right),  \label{eq:kij}
\ee
while the line stress-energy tensor in terms of the line energy
density $\sigma$ and line tension $\vartheta$ is
\be
S^{i}{}_{j} = \left(
\begin{array}{rrr}

        -\sigma & 0  \\
           0    & -\vartheta

\end{array} \right).  \label{eq:sij}
\ee
We see from (\ref{eq:linese}) that the field equations become
\be
\sigma = -\frac{c^4}{8\pi G}{\cal K}^{\theta}{}_{\theta}~~~\mbox{and}~~~
\vartheta = -\frac{c^4}{8\pi G}{\cal K}^{\tau}{}_{\tau}. \label{eq:ksigma}
\ee

Considering the positive side of the top boundary [c.f.~(\ref{eq:jfup})] we
note that $\partial r/\partial {\eta}=\sqrt{\Lambda r^2-M}$ and
$\partial r/\partial {\eta}=-\sqrt{\Lambda r^2-M(r)}$ being
assigned to the negative side of the top boundary.  Armed
with (\ref{eq:twoform}), (\ref{eq:disc}), and (\ref{eq:ksigma}), and letting
$r=a$ at the boundary, we readily get
\be
\sigma = -\frac{c^4}{8\pi Ga}\left(\sqrt{\Lambda a^2-M}+\sqrt{\Lambda
a^2-M(a)}\right),
\ee
\be
\vartheta = -\frac{c^4}{8\pi G}\left(\frac{\Lambda a}{\sqrt{\Lambda a^2-M}}
+\Phi '(a)\sqrt{\Lambda a^2-M(a)}\right).
\ee

As expected, the energy density is negative at the throat of the
wormhole.  This simply implies that we have exotic matter, as previously
discussed.  The line tension is also negative, which implies that there
is a line pressure as opposed to a line tension.  This was
also expected as a pressure would be needed to prevent the collapse of the
wormhole throat.

A dynamic analysis of this solution can be obtained by letting the radius be
a function of time $a \mapsto a(t)$.  This method closely parallels that of
\cite{kn:Visser}.  Let the throat by described by $x^{\mu}(t,\phi)
= (t,a(t),\phi)$.  For the top boundary we get the three-velocity of a piece
of stress energy at the throat given by:
\be
U^{\mu}=\left(\frac{dt}{d\tau},\frac{da}{d\tau},0\right)=\left(\frac{\sqrt
{\Lambda a^2-M +\dot{a}^2}}{\Lambda a^2-M},\dot{a},0 \right).
\ee
The unit normal to the boundary can be calculated from the conditions
$U^{\mu}\xi_{\mu}=0$ and $\xi^{\mu}\xi_{\mu}=1$. The result is
\be
\xi^{\mu}=\left(\frac{\dot{a}}{\Lambda a^2-M},\sqrt{\Lambda a^2-M+\dot{a}^2},
0\right).
\ee
The $\phi \phi$ component is easily calculated from:
\be
K^{\phi}{}_{\phi}= \frac{1}{r} \left.\frac{\partial r}{\partial \eta}
\right|_{r=a},
\ee
giving
\be
K^{\phi}{}_{\phi}^{+} = \frac{\sqrt{\Lambda a^2-M+\dot{a}^2}}{a}.
\ee
The $\tau\tau$ component is more difficult using brute force.  Another
method is as follows.  First note that (see \cite{kn:Visser})
\be
K^{\tau}{}_{\tau} = \xi_{\mu}(U^{\mu}{}_{;\nu}U^{\nu})=
\xi_{\mu}A^{\mu}, \label{eq:ktta}
\ee
where $A^{\mu}$ is the three-acceleration of the throat.  Now $A^{\mu}\equiv
A\xi^{\mu}$ by circular symmetry so that, $K^{\tau}{}_{\tau} = A\equiv$
magnitude of the three-acceleration.  $A$ is determined by using the Killing
vector $k^{\mu}\equiv (\partial / \partial t)^{\mu} = (1,0,0)$ for the
underlying geometry.  In addition $k_{\mu}=(-(\Lambda a^2-M),0,0)$, and
\be
k_{\mu}\xi^{\mu} = -\dot{a},
\ee
\be
k_{\mu}U^{\mu} = -\sqrt{\Lambda a^2-M+\dot{a}^2}. \label{eq:kuup}
\ee
Comparing
\be
\frac{D}{D\tau}(k_{\mu}U^{\mu}) = k_{\mu;\nu}U^{\nu}U^{\mu} + k_{\mu}
\frac{DU^{\mu}}{D\tau} = -A\dot{a},
\ee
to the actual derivative of (\ref{eq:kuup}),
\be
\frac{D}{D\tau}(k_{\mu}U^{\mu}) = -\dot{a}\frac{a+\ddot{a}}{\sqrt{\Lambda
a^2-M+\dot{a}^2}},
\ee
we finally get
\be
K^{\tau}{}_{\tau}^{+} = \frac{\Lambda a + \ddot{a}}{\sqrt{\Lambda a^2 - M
+ \dot{a}^2}}.
\ee

Similar calculations for the middle section (\ref{eq:metric1}) give:
\be
K^{\phi}{}_{\phi}^{-} = -\frac{\sqrt{\Lambda a^2-M(a)+\dot{a}^2}}{a},
\ee
\begin{eqnarray}
K^{\tau}{}_{\tau}^{-} & = & \Phi '\sqrt{\Lambda a^2-M(a)+\dot{a}^2}
+\frac{\ddot{a}}{\sqrt{\Lambda a^2-M(a)+\dot{a}^2}} \nonumber  \\
                      &   & -\frac{\dot{a}^2(2\Lambda a - M'(a))}
{2\sqrt{\Lambda a^2-M(a)+\dot{a}^2}(\Lambda a^2-M(a))}.
\end{eqnarray}
One need only substitute into equations (\ref{eq:disc}) and (\ref{eq:ksigma})
to get the time dependent field equations
\be
\sigma = -\frac{c^4}{8\pi Ga}\left[\sqrt{\Lambda a^2-M+
\dot{a}^2}+\sqrt{\Lambda a^2-M(a)+\dot{a}^2}\right], \label{eq:dynsigma}
\ee
\small
\begin{eqnarray}
\vartheta & = & -\frac{c^4}{8\pi Ga}\left[\frac{\Lambda a + \ddot{a}}
{\sqrt{\Lambda a^2 - M+\dot{a}^2}}-\Phi '\sqrt{\Lambda a^2-M(a)+
\dot{a}^2}-\frac{\ddot{a}}{\sqrt{\Lambda a^2-M(a)+\dot{a}^2}} \right. \nonumber
\\
          &   & \left. +\frac{\dot{a}^2(2\Lambda a-M'(a))}{2\sqrt{\Lambda
a^2-M(a)+\dot{a}^2}(\Lambda a^2-M(a))}\right] . \label{eq:dynvtheta}
\end{eqnarray}
\normalsize

For a constant $\Phi (a)$ such that $\Phi '(a) = 0$, the conservation of
stress-energy implies \cite{kn:Visser},
\be
\dot{\sigma} = -\frac{\dot{a}}{a}(\sigma - \vartheta). \label{eq:sigmadot}
\ee
Introducing the length of the stress-energy ring $\ell = 2\pi a$, it is
easily seen that (\ref{eq:sigmadot}) may be rewritten in the more recognizable
form,
\be
\frac{D}{D\tau}(\ell\sigma) = \vartheta \frac{D}{D\tau}(\ell).
\ee

Assuming the equation of state $\sigma = \vartheta$ holds for
time dependent wormholes, we immediately have $\sigma=$\,constant
from (\ref{eq:sigmadot}).  Rearranging (\ref{eq:dynsigma}) and using
geometrodynamic units $c\equiv 1\equiv G$ we get the unsimplified differential
equation,
\be
\sqrt{\Lambda a^2-M+\dot{a}^2}+\sqrt{\Lambda a^2 -M(a) + \dot{a}^2} =
-8\pi \sigma a. \label{eq:sigmade}
\ee
This equation is very difficult to solve for most functions $M(a)$.  The
simplest case where $M(a)=M$ is the only one considered here.
Equation (\ref{eq:sigmade}) then becomes after some manipulation,
\be
\dot{a}^2 -a^2 (16\pi ^2 \sigma ^2 - \Lambda)-M = 0. \label{eq:sigmade1}
\ee
There are two cases to consider; the first one is where $\Lambda > 16\pi ^2
\sigma ^2$, and the second case is the opposite ($\Lambda < 16\pi ^2
\sigma ^2$).  For case 1, equation (\ref{eq:sigmade1}) has a cosine solution
as follows:
\be
a(\tau)=\frac{\sqrt{M}}{\sqrt{\Lambda - 16\pi ^2\sigma ^2}}\cos{\sqrt{\Lambda
- (16\pi ^2\sigma ^2)}\tau}.
\ee
This is a stable oscillatory solution.  In addition, the idea of a large
$\Lambda$ has already been introduced and conceivably possible using
relativistic bubbles.

For the second case we set $a=r_0$ at $\tau=0$, where $r_0$ is the initial
radius of the wormhole to get the solution,
\be
a(\tau) = \frac{1}{-2\sqrt{16\pi ^2\sigma ^2-\Lambda}}\left[ Q_+ e^{\pm
\sqrt{16\pi^ 2\sigma ^2 - \Lambda}\tau} + Q_-
e^{\mp\sqrt{16\pi ^2\sigma ^2 -\Lambda}\tau}\right],
\ee
with $Q_{\pm} \equiv (\sqrt{16\pi ^2\sigma ^2 - \Lambda})r_0 \pm
\sqrt{(16\pi ^2\sigma^2 -\Lambda) r_0^2 - 4M}$.  We see that as
$\tau\rightarrow\infty$, $a\rightarrow\infty$.  That is, the solution is
unstable to explosion.  However, for no $\tau$ does $a$ go to zero.  Hence
the solution is stable to collapse.  This is in agreement to the similar
solution for the $(2+1)$ case without a cosmological constant
\cite{kn:PerryMann}.  A similar approach can be used to analyse
the $(3+1)$ dimensional case.
%
%%%%%%%%%%%%%%%%%%%%%%%%%%%%%%%%%%%%%%%%%%%%%%%%%%%%%%%%%%%%%%%%%%%%%%%%%%%%%
%
\section{Discussion}

Inclusion of a cosmological constant modifies to some extent
the structure of a wormhole, permitting, for example, positive $\rho$ in
$(2+1)$ dimensions.  The zero-tidal force solution gave limits on the velocity
and showed a possibly very quick transit time.  However, these results were
only valid for relatively small velocities.  Additionally, this type of
solution needed an extremely large amount of exotic material.  Using a
bubble to modify the constant could solve this problem for the $(2+1)$
dimensional case but would create
the additional concern of the discontinuous boundary at the bubble surface.
The wormhole constructed by surgically grafting two solutions devoid of
matter around the general wormhole manifold gave negative line energy
density and line tension.  The addition of the cosmological constant to the
field equations allowed for a stable solution not found in previous papers.

The possibility for all of the above mentioned wormhole  universes hinge on
the existence of exotic matter.  Although no macroscopic exotic matter is
known to exist, quantum theory shows tantalizing hints that it may be
possible to manufacture exotic matter if it does not exist normally.

\section*{Acknowledgments}
This work was supported in part by the Natural Sciences and Engineering
Research Council of Canada.

\noindent
Note added: Upon completion of this manuscipt we became aware of a preprint
by S.W. Kim, H. Lee, S. Kim and J. Yang (``$(2+1)$ Dimensional Schwarzchild
de Sitter Wormhole'' SNUTP-93-51) in which $(2+1)$ dimensional
wormholes in de-Sitter space were analyzed.
%
%%%%%%%%%%%%%%%%%%%%%%%%%%%%%%%%%%%%%%%%%%%%%%%%%%%%%%%%%%%%%%%%%%%%%%%%%%%%
%

\end{document}